# AN INVESTIGATION ON FUZZY LOGIC CONTROLLERS (TAKAGI-SUGENO & MAMDANI) IN INVERSE PENDULUM SYSTEM

Qasem Abdollah Nezhad [1], Javad Palizvan Zand [2] and Samira Shah Hoseini [3]

[1&3]Department of Mechatronics Engineering, Science and Research Branch Kurdistan Islamic Azad University, Tabriz, Iran
qasem_a_nejad@yahoo.com
smr.shhoseini@gmail.com

[2]Department of Civil Engineering, Tabriz University, Tabriz, Iran
j.palizvan@gmail.com

## ABSTRACT

*The concept of controlling non-linear systems is one the significant fields in scientific researches for the purpose of which intelligent approaches can provide desirable applicability. Fuzzy systems are systems with ambiguous definition and fuzzy control is an especial type of non-linear control. Inverse pendulum system is one the most widely popular non-linear systems which is known for its specific characteristics such as being intrinsically non-linear and unsteady. Therefore, a controller is required for maintaining stability of the system Present study tries to compare the obtained results from designing fuzzy intelligent controllers in similar conditions and also identify the appropriate controller for holding the inverse pendulum in vertical position on the cart.*

## KEYWORDS

*Fuzzy systems, Inverse Pendulum, intelligent controllers, non-linear systems.*

## 1. INTRODUCTION

By introducing fuzzy sets and the concept of membership degree Dr. Lotfali Asgar Zadeh founded fuzzy logic in 1965. The concept of fuzzy logic can be used for two different meanings. Considering its limited meaning, fuzzy logic is a logic system aimed at stimulating the approximate reasoning and within this domain it is considered as an extension of multiple values logic and real numbers within [0, 1] are verity values of this logic. In terms of wide meaning it equals to the theory of fuzzy sets [2, 3].
The fuzzy sets theory can be defined as a proposition for acting in uncertain circumstances. This theory is able to give a mathematical configuration for unclear concepts, variables and systems and also can allow reasoning, inference, control and deciding in uncertain conditions.
Fuzzy systems are dependent on knowledge or principles. The core of fuzzy system is a data base which is composed of ((if-then)) principles. A fuzzy ((if-then)) principle is an if-then expression some words of which are determined by dependency functions. The engine of fuzzy inference combines these principles into a registration of fuzzy sets in input space to fuzzy sets in output space. Fuzzy logic is used for designing expert systems which simulate real world [5].
In this study we will examine the outcomes from designing intelligent controllers of Takagi-sugeno and Mamdani type in similar conditions. Then, on the basis of these results and comparing them, we will identify the most appropriate controller which can hold the pendulum in vertical position on cart with more sensitivity and accuracy. These results are consequences of system response to sinusoidal and square inputs.

## 2. Inverse Pendulum System

Inverse pendulum system is considered as a controlled system as depicted below [6]:

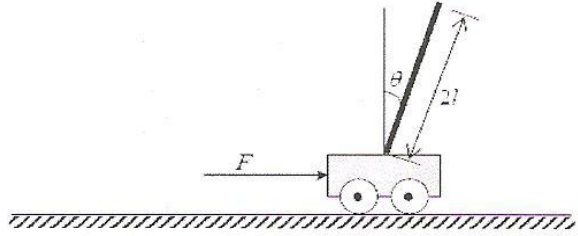

Figure 1. Inverse pendulum system and controlling input

As can be seen in the above figure, F represents controlling input, l is half the length of inverse pendulum, and θ indicates the angle at which the inverse pendulum deviates from perpendicular position. Inverse pendulum system dynamics can be presented as follow:

$$\ddot{\theta} = \frac{g\sin\theta - \dfrac{\cos\theta(F + ml\dot{\theta}^2 \sin\theta)}{m+M}}{l\left(\dfrac{4}{3} - \dfrac{m\cos^2\theta}{m+M}\right)}$$

$$\ddot{x} = \frac{F + ml(\dot{\theta}^2 \sin\theta - \ddot{\theta}\cos\theta)}{m+M} \quad (1)$$

where, g is the constant of gravity which equals $10 \, m/s^2$ is pendulum mass, and M is cart mass, and the parameters of the model are considered as: $l = 1^m$, $m = 0.1^{kg}$, $M = 1^{kg}$

It should be noted that in designing fuzzy intelligent controllers (sections 3 and 4) inverse pendulum system is controlled by four variables of state, namely horizontal distance $x$, horizontal velocity $\dot{x}$, deviation angle $\theta$, and angular velocity of deviation $\dot{\theta}$. These variables are in the range of the following limits:

$$-0.3 \prec \theta \prec 0.3$$
$$-1 \prec \dot{\theta} \prec 1$$
$$-3 \prec x \prec 3 \quad (2)$$
$$-6 \prec \dot{x} \prec 6$$

and besides, the initial conditions are assumed to be zero [7].

$$x_0 = \begin{bmatrix} 0 & 0 & 0 & 0 \end{bmatrix} \quad (3)$$

## 3. Designing a fuzzy logic controller of Takagi-sugeno type

The fuzzy controllers are designed by using the graphical interface of fuzzy logic toolbox. The fuzzy deduction system (FIS) diagram block of Takagi-sugeno type is illustrated as follows:

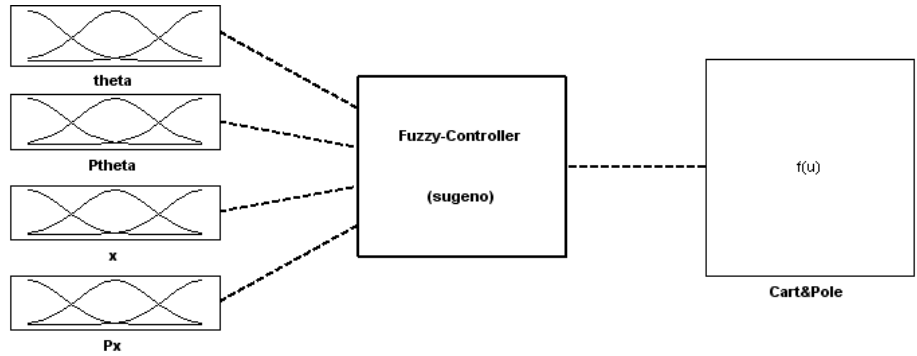

Figure 2. Inverse pendulum system controller

Four dependency functions have been defined for each input; hence 16 deduction rules will be required:

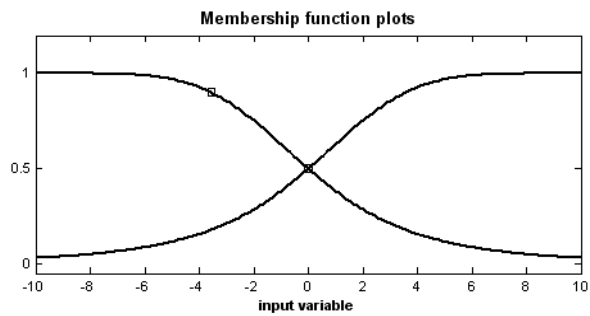

Figure 3. Dependency functions of Inputs

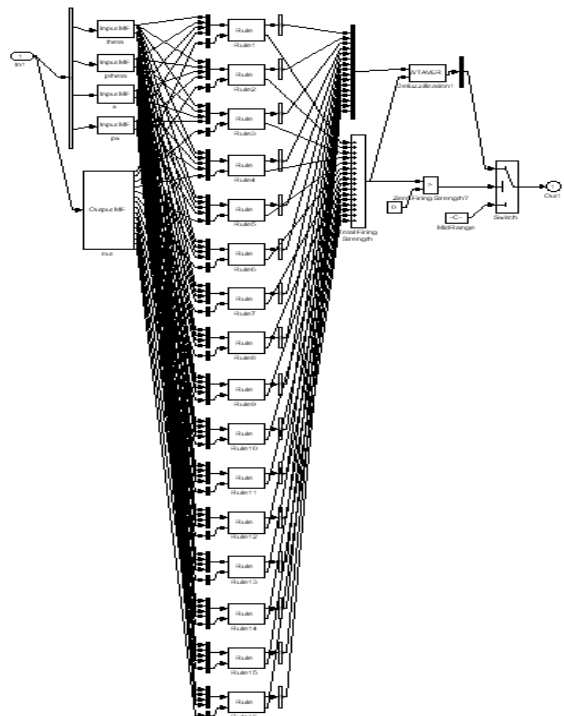

Figure 4. Principles of fuzzy inference

In figure 4, the number of blocks is equal to number of rolls and input and output lines to and from these blocks are fuzzy rules and relations. The input and output variables are depicted on the left and right side of the figure respectively.

Inputs dependency functions are defined as Gauss type which are depicted in figure 3 and a linear composition of inputs is applied on output [8, 9, and 10].

## 4. Designing a fuzzy logic controller of Mamdani type

The following figures show the fuzzy deduction system (FIS) diagram block of Mamdani type:

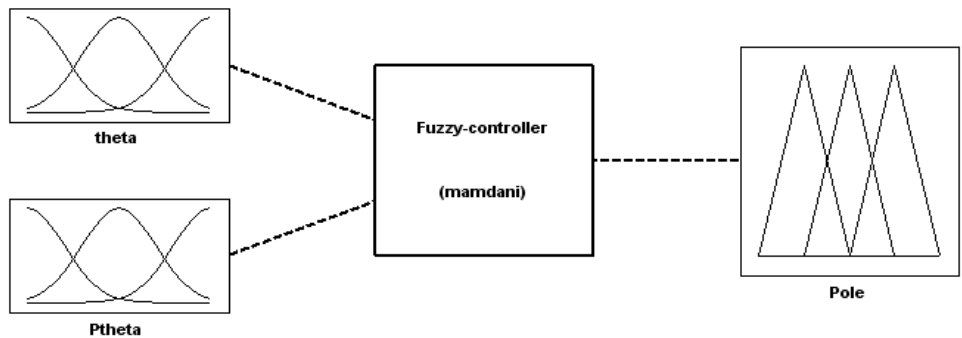

Figure 5. Pole position controller

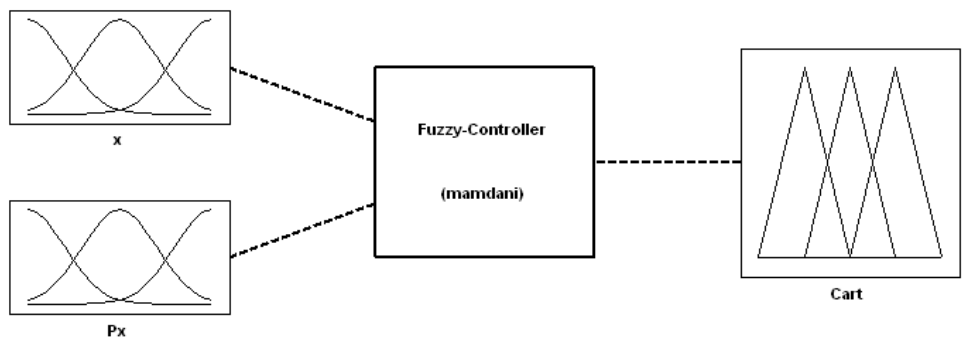

Figure 6. Cart position controller

Two inputs for each controller and totally twelve dependency functions for these two inputs have been defined, hence 35 deduction rules will be required [8, 9 and10].

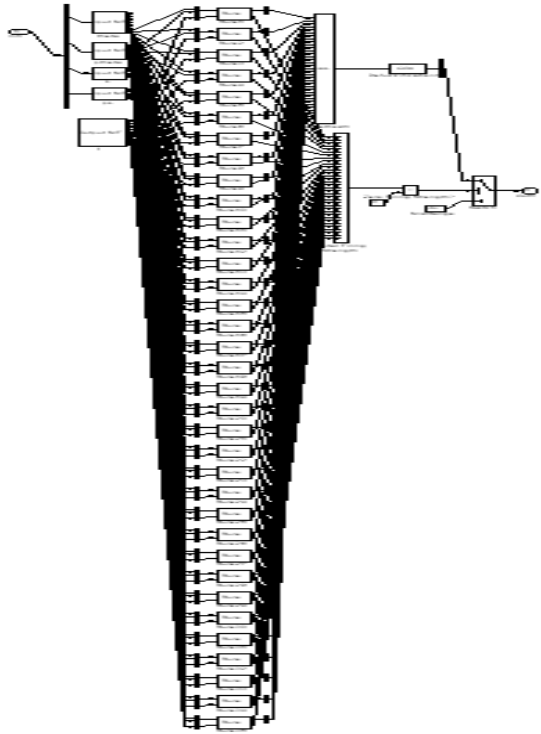

Figure 7. Principles of fuzzy inference

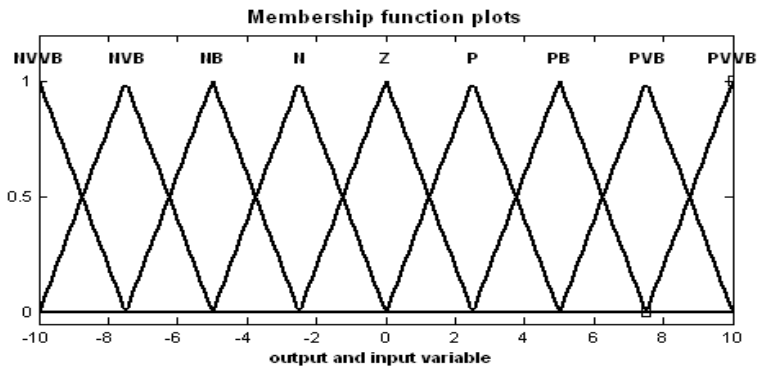

Figure 8. Dependency functions of input and output

The dependency functions of input and output are defined as triangular type which can be seen in the figure below. These functions are expressed in verbal terms as (NVVB), (NVB), (NB), (N), (Z), (P), (PB), (PVB), and (PVVB). The scope of inputs for both controllers (Tkagi-sugeno and Mamdani) has obtained on the basis of trial and error approach. The utilized fuzzy rules in both controllers have also obtained based on the type of reaction shown by the inverted pendulum and also through experiments and repetitive examinations [4].

## 5. Evaluating the output results of the intelligent controllers in simulation (system response to square input)

If the applied force is of square type, the fuzzy intelligent controllers output will track the input in a shape similar to the curves below. Simulation time has been assumed to be ten seconds for both controllers. Figures 9, 10, 11, 12, 13 are related to Takagi-sunego type controller and figures 14, 15, 16, 17, 18 are presentations of Mamdani type controller.

In the following illustration, the highest level of deviation from system input state on positive area and at the moment of 9.8146 seconds equals to 1.125 Newton.

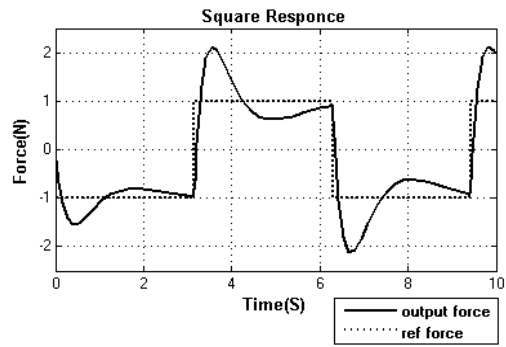

Figure 9. Response to sinusoidal input

In the following illustration, the highest deviation of pendulum position from balanced state (zero point) on negative area and at the moment of 6.7865 seconds equals to 0.063 Radians.

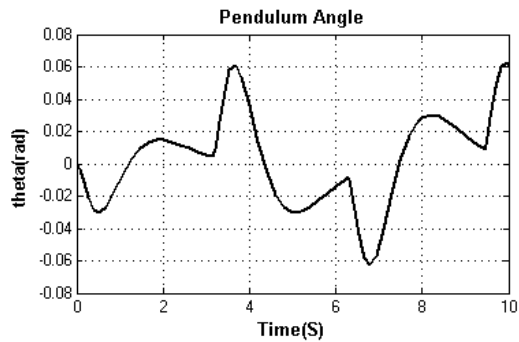

Figure 10. Pendulum angle

In the following illustration, the highest deviation of pendulum velocity from balanced state (zero point) on positive area and at the moment of 3.3 seconds equals to 0.1882 Radians per second.

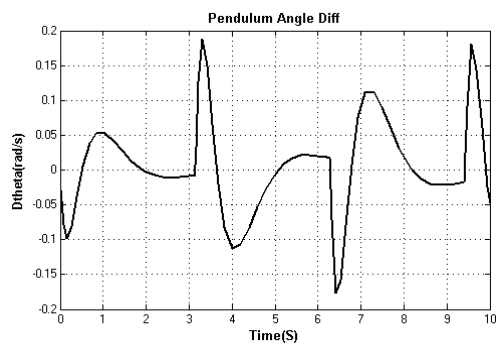

Figure 11. Pendulum angle derived

In the following illustration, the highest deviation of cart position from balanced state (zero point) on negative area and at the moment of 9.6911 seconds equals to 0.3651 meters.

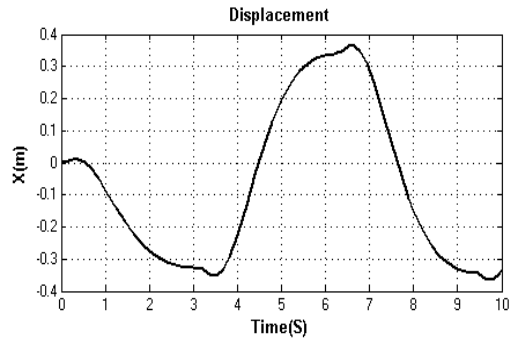

Figure 12. Displacement

In the following illustration, the highest deviation of cart velocity from balanced state (zero point) on negative area and at the moment of 7.4928 seconds equals to 0.4804 meters per second.

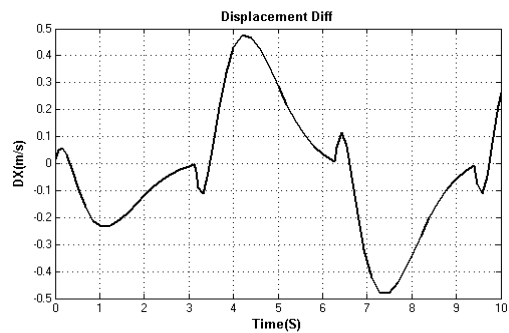

Figure 13. Derivative of displacement

In the following illustration, the highest level of deviation from system input state on negative area and at the moment of 7.2671 seconds equals to 0.9321 Newton.

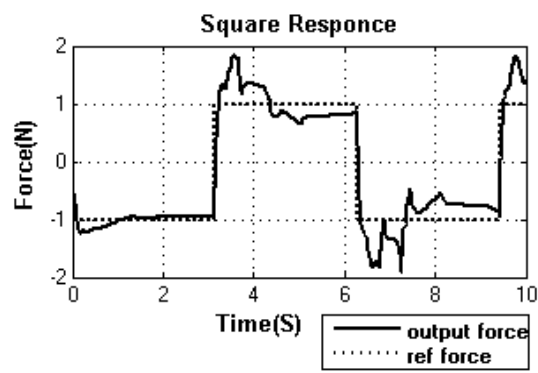

Figure 14. Response to square input

In the following illustration, the highest deviation of pendulum position from balanced state (zero point) on negative area and at the moment of 6.7014 seconds equals to 0.0391 Radians.

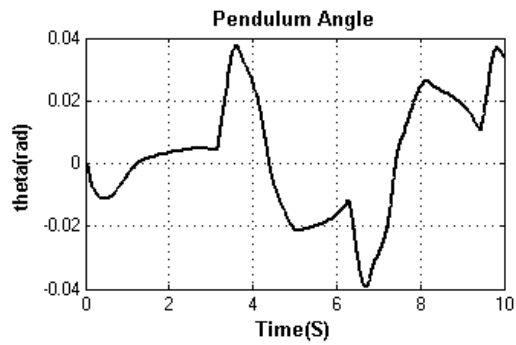

Figure 15. Pendulum angle

In the following illustration, the highest deviation of pendulum velocity from balanced state (zero point) on negative area and at the moment of 7.4 seconds equals to 0.1136 Radians per second.

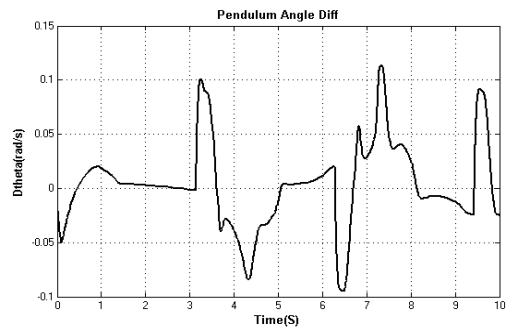

Figure 16. Pendulum angle derived

In the following illustration, the highest deviation of cart position from balanced state (zero point) on negative area and at the moment of 9 seconds equals to 0.2711 meters.

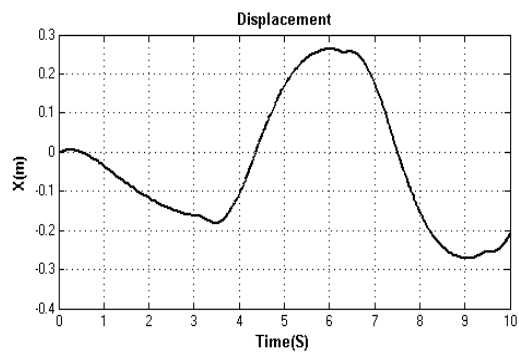

Figure 17. Displacement

In the following illustration, the highest deviation of cart velocity from balanced state (zero point) on negative area and at the moment of 7.3619 seconds equals to 0.3936 meters per second.

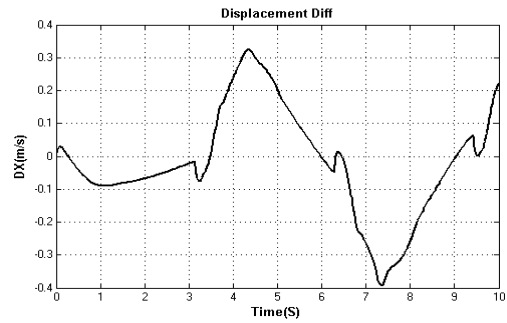

Figure 18. Derivative of displacement

## 6. Evaluating the output results of the intelligent controllers in simulation (system response to sinusoidal input)

If the applied force is of sinusoidal type, the fuzzy intelligent controllers output will track the input in a shape similar to the curves below. Simulation time has been assumed to be ten seconds for both controllers. Figures 19, 20, 21, 22, 23 are related to Takagi-sunego type controller and figures 24, 25, 26, 27, 28 are presentations of Mamdani type controller.

In the following illustration, the highest level of deviation from system input state on positive area and at the moment of 1.3858 seconds equals to 0.1537 Newton.

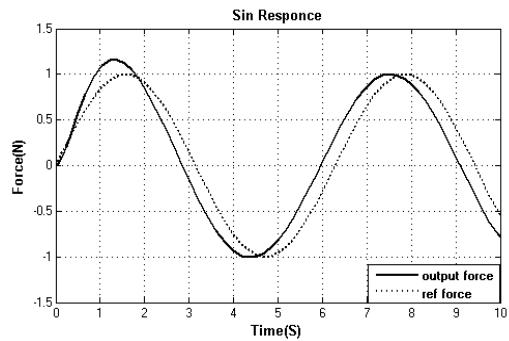

Figure 19. Response to sinusoidal input

In the following illustration, the highest deviation of pendulum position from balanced state (zero point) is on two areas, one on positive area and at the moment of 6.1844 seconds and the other on negative area and at the moment of 9.3489 which equals to 0.0283 Radians.

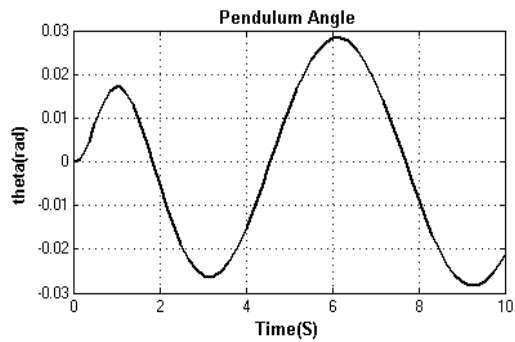

Figure 20. Pendulum angle

In the following illustration, the highest deviation of pendulum velocity from balanced state (zero point) on negative area and at the moment of 1.9 seconds equals to 0.0325 Radians per second.

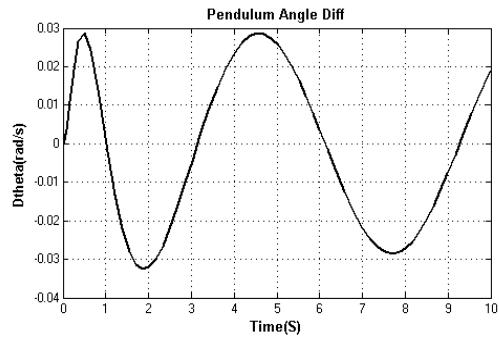

Figure 21. Pendulum angle derived

In the following illustration, the highest deviation of cart position from balanced state (zero point) on negative area and at the moment of 6.1844 seconds equals to 0.2968 meters.

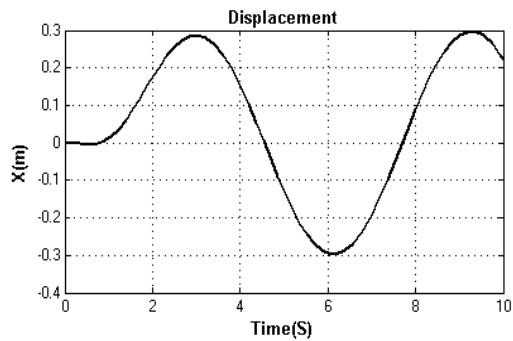

Figure 22. Displacement

In the following illustration, the highest deviation of cart velocity from balanced state (zero point) on positive area and at the moment of 7.7044 seconds equals to 0.2976 meters per second.

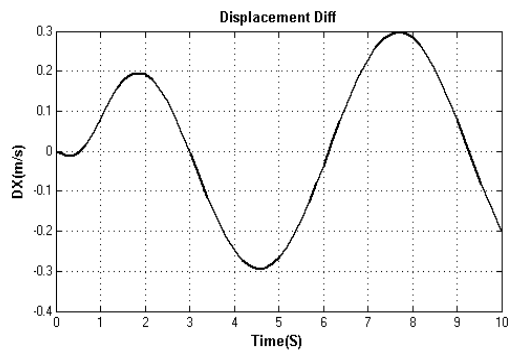

Figure 23. Derivative of displacement

In the following illustration, the highest level of deviation from system input state on positive area and at the moment of 1.4799 seconds equals to 0.0585 Newton.

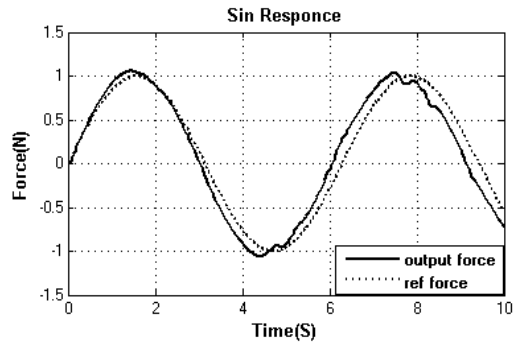

Figure 24. Response to sinusoidal input

In the following illustration, the highest deviation of pendulum position from balanced state (zero point) on negative area and at the moment of 9.2717 seconds equals to 0.0237 Radians.

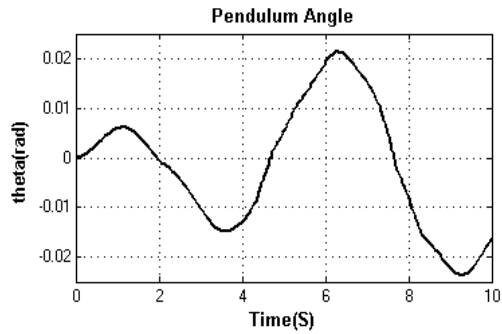

Figure 25. Pendulum angle

In the following illustration, the highest deviation of pendulum velocity from balanced state (zero point) on negative area and at the moment of 7.6 seconds equals to 0.0346 Radians per second.

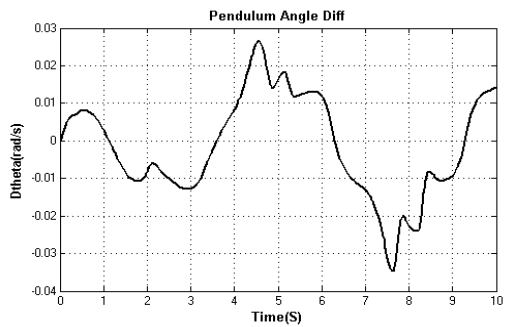

Figure 26. Pendulum angle derived

In the following illustration, the highest deviation of cart position from balanced state (zero point) on positive area and at the moment of 9.2 seconds equals to 0.2375 meters.

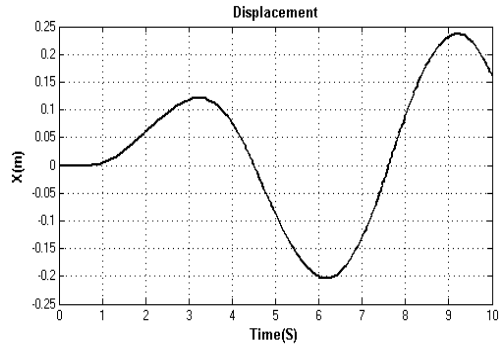

Figure 27. Displacement

In the following illustration, the highest deviation of cart velocity from balanced state (zero point) on positive area and at the moment of 7.6361 seconds equals to 0.2414 meters per second.

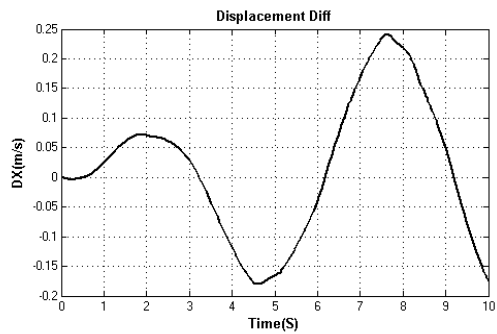

Figure 28. Derivative of displacement

## 7. CONCLUSIONS

A wide range of applications dealing with controlling complicated and non-linear systems have focused on fuzzy control systems in recent years. Although Takagi-sugeno type fuzzy control systems are effective in controlling non-linear systems, unlike Mamdani type fuzzy control systems all of the system equations are required to be completely known for designing Tkagi-sugeno controllers [1].

Considering the results mentioned in section 5 which are related to system response to square input, it can be easily realized that system over shoot in takagi-sugeno type controller (1.125) is more –with a difference about 0.193– than that of Mamdani type (0.932). Thus assuming equal experimental conditions, the most appropriate controller for square input would be Mamdani type. Also considering the results mentioned in section 6 which are related to system response to sinusoidal input, it can be easily realized that system over shoot in takagi-sugeno type controller (0.1537) is more -with a difference about 0.0952- than that of Mamdani type (0.0585). Thus assuming equal experimental conditions, the most appropriate controller for sinusoidal input would be Mamdani type.

In both sections (5, 6) considering to figures which get from results, displacement and velocity of the cart with deviation and angular velocity of the pendulum Mamdani type controller is less than that of takagi-sugeno type.

Finally, we can draw this conclusion from sections 5 and 6 that Mamdani type fuzzy smart controller possessing lower over shoot and higher accuracy and sensitivity can effectively hold the reverse pendulum in upright position on the cart. Therefore, providing similar experimental conditions, the suitable controller for reverse pendulum system will be Mamdani type fuzzy smart controller.

**Authors**

**Qasem Abdollah Nezhad** was born in 1984, He received bachelor of Science in Mechanical Engineering from the Department of Mechanical Engineering, Islamic Azad University, Tabriz branch, Iran, in 2011, and Master of science Mechatronics Engineering from the Department of Mechatronics Engineering, Science and Research Islamic Azad University, Kurdistan branch, Iran, in 2013.Currently he is a lecturer, He is the author of several papers in National Conference, His research interests includes robotics, mechatronics, image processing, control systems, fuzzy control application.

**Javad Palizvan Zand** was born in 1986, He received bachelor of science in Civil Engineering from the Department of Civil Engineering, Islamic Azad University, Tabriz branch, Iran, in 2009, and Master of science Structural Engineering from the Department of Civil Engineering, Tabriz University, Iran, in 2012, He is a member of the Young Researchers and Elite Club, Tabriz Branch, Islamic Azad University, Currently he is a lecturer, He is the author of several papers in National Conference.

**Samira Shah Hoseini** was born in 1982, She received bachelor of Science in Electrical and Communication Engineering from the Department of Tabriz University, Iran, in 2007, and Master of science Mechatronics Engineering from the Department of Mechatronics Engineering, Science and Research Islamic Azad University, Kurdistan branch, Iran, in 2013.Currently she is a lecturer, She is the author of several papers in National Conference, Her research interests includes robotics, mechatronics, image processing, control systems, fuzzy control application.